\documentclass[12pt]{article}  
\usepackage{amsmath,amssymb,bm}
\usepackage{epsfig}
\usepackage{graphicx}
\usepackage{amsfonts}
\usepackage{epstopdf}
\usepackage[usenames,dvipsnames]{color} 
\usepackage{float}

\newcommand{\beqa}{\begin{eqnarray}}
\newcommand{\eeqa}{\end{eqnarray}}
\newcommand{\bd}[1]{ \mbox{\boldmath $#1$}}

\begin{document}
\def\ii{\'\i}

\title{The Hydrogen Atom within a pseudo-complex Qunatum Mechanics, 
involving a minimal length}

\author{
P. O. Hess$^{1,2}$ \\
{\small\it
$^1$ Instituto de Ciencias Nucleares, Universidad Nacional 
Aut\'onoma de M\'exico,}\\
{\small\it
Circuito Exterior, C.U., 
A.P. 70-543, 04510 M\'exico D.F., Mexico}\\
{\small\it
$^2$ Frankfurt Institute for Advanced Studies, Johann Wolfgang 
Goethe Universit\"at,}\\
{\small\it
Ruth-Moufang-Str. 1, 60438 Frankfurt am Main, Germany}

}

\maketitle

\abstract{
The hydrogen atom is investigated, 
within a pseudo-complex extension of the coordinates and momenta,
which introduces a minimal length scale ($l$) and 
results into a non-commutative Quantum Mechanics. After
resuming the
pseudo-complex extension of Quantum Mechanics,
the modified energies of the hydrogen atom are
deduced, producing corrections of the 
order of the square of
the minimal length scale. Using 
the Lamb Shift, we obtain an upper boundary for
the minimal length scale $l$, orders of
magnitude more restrictive than former estimations.
}

\section{Introduction}
\label{intro}

At small distances, for example the Planck length
$l_p=10^{-33}$cm, one expects that Quantum Mechanics is 
affected by a minimal length scale $l>l_p$, i.e., that
the commutation relations between coordinates and momenta
may change such that even coordinates do not commute.
To my knowledge, the first to propose a non-commutative
Quantum Mechanics was H. S. Snyder \cite{snyder1947}. 
Since then, many more contributions where published, where I
mention only the definition of the {\it Moyal product} 
\cite{moyal}. Other mainstream theories,
as the {\it String Theory} \cite{string} and 
{\it Quantum Loop Theory} \cite{loop}, have as an essential structure
a minimal length, either in terms of the dimension of a string
or the granulation of space-time. The resulting theory is
in general quite involved.

Any novel idea to ease the handling of a minimal length
scale is of importance and here I will propose such a new path:
In \cite{pcQM} a proposal was presented on how to
obtain a non-commutative Quantum Mechanics. The
novelty consists in using a standard quantization rule
in an 8-dimensional space, extending algebraically 
the coordinates to so-called 
{\it pseudo-complex coordinates} (see Section 
\ref{pcQM}), doubling the dimension of the space to 8.
The physical 4-dimensional space is then represented by a
4-dimensional subspace within this larger space.
The attractive part of the pseudo-complex extension
is its simplicity, as it uses the standard quantization rule
in the 8-dimensional space, and the introduction of an effective
length scale {\it parameter}, 
simulating a granulation of space,
{\it without having to give up Lorentz symmetry
in the larger space}. 

In the present contribution the
Hydrogen Atom is investigated within the pseudo-complex extension of 
Quantum Mechanics \cite{pcQM} and an upper limit of the
minimal length scale is obtained. In Section \ref{pcQM}
the pseudo-complex Quantum Mechanics in a flat space 
will be resumed and in Section \ref{HyAt} this theory
will be applied to the Hydrogen atom. In Section
\ref{concl} Conclusions will be drawn.

For a better understanding of the results,
in the Appendix we resume the dimensions in natural units 
($\hbar=c=1$) of the variables,
parameters and operators used.

\section{Pseudo-complex Quantum Mechanics of the Hydrogen Atom}
\label{pcQM}

The dynamical symmetry group of the hydrogen atom is $SO(4)$, whose 
generators are the sum of the three angular momentum and three 
Runge-Lenz vectors \cite{Greiner-Muller}
in a three dimensional space, depending on the coordinates $x_i$ and momenta $p_j$
($i,j=1,2,3$). As shown in \cite{Greiner-Muller} these six operators can
be redefined in terms of an operator ${\bd L}_{ij}$ where now the indices
run from 1 to 4. This operator is represented in terms of a combination of 
four abstract coordinates 
and their conjugate momenta in a 4-dimensional space and an example will be
given further below. This has to be extended such that the theory contains a minimal length.

In \cite{pcQM} a proposal was presented for a quantization
in flat Cartesian space with a minimal length parameter. The
main ingredient is the definition of pseudo-complex
coordinates and momenta ($i,j,k = 1,2,3,4$, being
the indices for the Cartesian coordinate in an abstract 4-dimensional space
\cite{Greiner-Muller})
\beqa
X_i & = & x_i + I l y_i
~.~ P_i ~=~ p^x_i + Il p^y_i
\nonumber \\
X_i & = & X^+_i \sigma_+ + X^-_i \sigma_-
~,~ P_i ~=~ P^+_i \sigma_+ + P^-_i \sigma_-
~~~,
\label{QM-1}
\eeqa
with $\sigma_\pm = \frac{1}{2}\, \left( 1 \pm I \right)$
where in addition an alternative representation
of the coordinates in terms of the zero-divisor components
$X_i^{\pm}$ ($P_i^{\pm}$)
is given. Due to the property of
$\sigma_\pm^2=1$ and $\sigma_+\sigma_-=0$ the coordinates
(momenta) imply a so-called {\it zero divisor}.
The coordinates $x_i$ and $y_i$ form an 8-dimensional
space. The operator $I$ has the property $I^2=1$, which is
the reason to call the coordinates and momenta 
{\it pseudo-complex} \cite{PPNP}. 
The standard definition and meaning of the four coordinates is explained
in \cite{Greiner-Muller}.
The reason to use pseudo-complex coordinates is explained in
\cite{PPNP,kelly}, where it is shown that only these coordinates
represent a viable extension from real ones, because otherwise
the theory contains ghost and/or tachyon solutions. 

In \cite{pcQM} is is shown that the pseudo-complex term of
the length element $d\omega^2$ = $g_{ij}dX_i dX_j$ is given by (in Cartesian components)
$\delta_{ij}dx_i dy_j$, which has to be 
set to zero in order that the length element is a real quantity and 
where $\delta_{ij}$ is the Cartesian metric. The solution is
$u_i=\frac{dx_i}{d\tau}$ \cite{PPNP}, where $\tau$
is the proper time. 
The dimension of $y_i$ in natural units is therefore 1.
A similar consideration can be done, as M. Born did \cite{PPNP},
defining in the momentum space a length element and in
analogy to the pseudo-imaginary part of the length element in the x-space 
and using (\ref{QM-1}) one obtains that
the units of $p_i^y$ are GeV$^2$. This is also the reason why
in (\ref{QM-1}) the pseudo-imaginary part has the factor of $l$, whose
unit is GeV$^{-1}$, such that the units of $P_i$ remains GeV.
Analogue to the coordinates the $p^y_i$ has to be the eigen-time derivative of 
$p^x_i$, i.e., $p^y_i=\frac{dp^x_i}{d\tau}$.

The condition that the length element $d\omega^2$ is real,
reduces again the 8-dimensional space to a 4-dimensional one.
Thus, the physical space is a 4-dimensional subsurface 
embedded in the 8-dimensional pseudo-complex space.

The coordinates $X_i$ ($X^\pm_i$) and momenta $P_i$
($P^\pm_i$) are quantized in the standard way, namely
\beqa
\left[X_i , P_j \right] & = & i \delta_{ij}
~\rightarrow~ \left[X^\pm_i , P^\pm_j \right] ~=~  
i \delta_{ij} ~,~
\left[ X^\pm_i , P^\mp_j  \right] ~=~ 0
~~~.
\label{QM-2}
\eeqa

Using \ref{QM-2}),
it is shown in \cite{pcQM} that the resulting commutation
relations between standard coordinates ($x_i$) and
momenta ($p_j$) are (here rewritten in terms
of Cartesian coordinates)
\beqa
\left[x_i , x_j \right] & = & 
-l^2\left[ y_i , y_j \right]
~,~
\left[ x_i , y_j \right] = - \left[y_i , x_j \right]
\nonumber \\
\left[p^x_i , p^x_j \right] & = & 
-l^2
\left[ p^y_i , p^y_j \right]
~,~
\left[ p^x_i , p^y_j \right] = 
- \left[p^y_i , p^x_j \right]
\nonumber \\
\left[x_i , p^x_j \right]
& = & i\hbar \delta_{ij}
 - 
l^2\left[y_i , p^y_j \right]
~,~
\left[ x_i , p^y_j \right] ~ = ~ 
- \left[y_i , p^x_j \right]
~~~.
\label{QM-3}
\eeqa
With the help of the Appendix it can be shown
that the dimensions are maintained.
The relations in (\ref{QM-3}) are similar to those obtained
by H. S. Snyder \cite{snyder1947}.

The generators of the pseudo-complex $SO(4)$ group,
$SO_{pc}(4)$, are given by ($i,j$ = 1, 2, 3, 4)
\beqa
{\bd L}_{ij} & = & X_i P_j - X_j P_i ~=~
{\bd L}^+_{ij} \sigma_+ + {\bd L}^-_{ij} \sigma_-
~,~
{\bd L}^\pm_{ij} ~ = ~ X^\pm_i P^\pm_j - X^\pm_j P^\pm_i
~~~.
\label{QM-4}
\eeqa
These pseudo-complex generators satisfy the commutation relations
\beqa
\left[ {\bd L}_{ij},{\bd L}_{kq}\right]
& = & 
i \left( \delta_{jk} {\bd L}_{q i}
+\delta_{q j} {\bd L}_{ik}
+\delta_{i k} {\bd L}_{jq}
+\delta_{i q} {\bd L}_{k j} \right)
\nonumber \\
\left[ {\bd L}^\pm_{ij},{\bd L}^\pm_{kq}\right]
& = & 
i \left( \delta_{j k} {\bd L}^\pm_{q i}
+\delta_{q j} {\bd L}^\pm_{i k}
+\delta_{i k} {\bd L}^\pm_{j q}
+\delta_{i q} {\bd L}^\pm_{k j} \right)
\nonumber \\
\left[ {\bd L}^+_{ij},{\bd L}^-_{kq} \right] & = & 0
~~~.
\label{QM-5}
\eeqa

Thus, the quantization rules are {\it the same} as in standard 
Quantum Mechanics, with the difference that it is now 
in an 8-dimensional space. 
However, this is no longer the case
when the real and pseudo-imaginary part of ${\bd L}_{ij}$ are considered
separately.
As can be seen by (\ref{QM-3}) the algebra in these
components are non-commutative \cite{pcQM}. In this way, the
non-commutative behavior of the coordinates 
is represented by the much
simpler commutation relations in the components of the
zero-divisor.

Before discussing the Hydrogen Atom in the pseudo-complex
formulation, we have to define various operators and relations:
Using (\ref{QM-1}) and (\ref{QM-4}), we express the
${\bd L}^\pm_{ij}$ in terms of operators which depend
on $x_i$, $y_j$ and their momenta, i.e.,
\beqa
{\bd L}^\pm_{ij} & = &
\left( {\bd L}^x_{ij} + l^2 {\bd L}^y_{ij}\right)
\pm l\left( {\bd L}^{xy}_{ij} + {\bd L}^{yx}_{ij}\right)
\nonumber \\
{\bd L}^x_{ij} & = & x_i p^x_j - x_j p^x_i
~,~
{\bd L}^y_{ij} ~ = ~ y_i p^y_j - y_j p^y_i
\nonumber \\
{\bd L}^{xy}_{ij} & = & x_i p^y_j - x_j p^y_i
~,~
{\bd L}^{yx}_{ij} ~ = ~ y_i p^x_j - y_j p^x_i
~~~.
\label{QM-6}
\eeqa
Though, for the operators the structure look like the generators of
$SO(4)$
groups, {\it they are not} \cite{pcQM}. The reason lies in the non-commutative
behavior of the coordinates and momenta. Using natural units (see the Appendix),
the units of the operators in (\ref{QM-6}) are 
$\left[{\bd L}^x_{ij} \right]=1$, $\left[{\bd L}^y_{ij} \right]={\rm GeV}^2$,
$\left[{\bd L}^{xy}_{ij} \right]={\rm GeV}$, $\left[{\bd L}^{yx}_{ij} \right]={\rm GeV}$.

The real (pseudo-imaginary) part is obtained by taking the sum
(difference) of the plus and minus zero-divisor component of the 
generators, which leads to
\beqa
{\bd L}^R_{ij} & = & \frac{1}{2}\left( {\bd L}^+_{ij} 
+ {\bd L}^-_{ij}
\right) ~=~ {\bd L}^x_{ij} + l^2 {\bd L}^y_{ij}
~,~
{\bd L}^I_{ij} ~ = ~ \frac{1}{2}\left( {\bd L}^+_{ij} 
- {\bd L}^-_{ij}
\right) ~=~ l 
\left( {\bd L}^{xy}_{ij} + {\bd L}^{yx}_{ij} \right)
~~~,
\label{QM-7}
\eeqa
where $R$ refers to the real and $I$ to the pseudo-imaginary 
part.
By construction the ${\bd L}^R_{ij}$ 
and ${\bd L}^I_{ij}$, the pseudo-real and pseudo-imaginary part
of ${\bd L}_{ij}$ = ${\bd L}^R_{ij}$ + ${\bd L}^I_{ij}$, form
an algebra $SO_R(4)$ and $SO_I(4)$ respectively. 
Note, that $\left[ {\bd L}^R\right]$ = 1 and 
$\left[ {\bd L}^I\right]$ = GeV. The same hold for the
operators $M^R_{ij}$ and $M^I_{ij}$.

Following \cite{Greiner-Muller}, one defines the operators
\beqa
{\bd L}^2 & = & {\bd L}_{1}^2 + {\bd L}_{2}^2 +
{\bd L}_{3}^2 
~ , ~
{\bd M}^2 ~ = ~ {\bd M}_{1}^2 + {\bd M}_{2}^2 +
{\bd M}_{3}^2 
~~~, 
\label{QM-8}
\eeqa
with ${\bd L}_1={\bd L}_{23}$, 
${\bd L}_2= {\bd L}_{13}$, 
${\bd L}_3={\bd L}_{12}$,
${\bd M}_1={\bd L}_{14}$, 
${\bd M}_2={\bd L}_{24}$, 
${\bd M}_3={\bd L}_{34}$.
The $SO_{pc}(4)$ has two pseudo-complex Casimir operators,
where we need only the relevant one for our purpose, namely
\cite{Greiner-Muller}
\beqa
{\bd C}^R & = & \frac{1}{2}\left( {\bd L}_R^2 
+ {\bd M}_R^2
\right) ~ \rightarrow ~ 2 k_R (k_R + 1)
~~~, 
\label{QM-9}
\eeqa
where on the right its eigenvalue is listed,
with $k_R$ is integer or half-integer.

\section{The Hydrogen Atom as an Example}
\label{HyAt}

According to \cite{Greiner-Muller}, the Hamiltonian of
the Hydrogen Atom is given by (in natural units)
\beqa
{\bd H} & = & - \frac{\mu e^2}
{4\left({\bd C}^x+\frac{1}{2}\right)}
~~~,
\label{HyAt-1}
\eeqa
with ${\bd C}^x$ the operator which for $l=0$ is a Casimir operator
of the $SO(4)$ group \cite{Greiner-Muller}.
The $e$ is the elementary charge and $\mu$ is the reduced mass
of the proton-electron system, i.e., approximately the electron mass
$m_e$. The units are $[\mu ]=$GeV, $[e^2 ]=1$.
The ${\bd C}^x$ can be expressed as
${\bd C}^x$ =  
$\frac{1}{2}\left( {\bd L}_x^2 + {\bd M}_x^2
\right)$.
For $l>0$ the ${\bd C}^x$ operator is in general {\it not} a Casimir
operator of a $SO(4)$ group.

In order to express the Hamiltonian in terms of the ${\bd L}$ 
and ${\bd M}$ operators, we use (\ref{QM-7}) 
and the same for the ${\bd M}^x_{ij}$ and ${\bd M}^R_{ij}$. 
I.e., we have
${\bd L}^x_{ij}={\bd L}^R_{ij} - 
l^2 {\bd L}^y_{ij}$
and
${\bd M}^x_{ij}={\bd M}^R_{ij} - 
l^2 {\bd M}^y_{ij}$. 
The ${\bd C}^x$ operator is written, up to the leading term
of corrections in the minimal length, as
\beqa
{\bd C}^x & \approx & {\bd C}^R - l^2 \sum_{\mu\nu}
\left[
\left({\bd L}^R \cdot {\bd L}^y\right)
+
\left({\bd M}^R \cdot {\bd M}^y\right)
\right]
~~~.
\label{HyAt-3}
\eeqa

Substituting this into (\ref{HyAt-1}) and expanding again
in leading order in $l^2$, we arrive at 
\beqa
{\bd H} & \approx & -\frac{\mu e^2}
{4 \left({\bd C}^R + \frac{1}{2}\right)}
\left( 1 + l^2
\frac{
\left[
\left({\bd L}^R \cdot {\bd L}^y\right)
+
\left({\bd M}^R \cdot {\bd M}^y\right)
\right]
}
{
4\left({\bd C}^R+\frac{1}{2}\right)
}
\right)
~~~.
\label{HyAt-4}
\eeqa
Note, that the units of $l^2$ is GeV$^{-2}$ and that the factor is of the units
GeV$^2$ (see section 2), i.e., the total unit of the product is 1.
What is not included here is
the correction due to the
Lamb shift. The correction calculated here is {\it only
due to the minimal length correction}, i.e., 
the $l$-dependent contribution should be 
smaller than the one of the Lambshift, otherwise
it should have been measured already.

The operator $4\left({\bd C}^R+\frac{1}{2}\right)$ in the
denominator has the eigenvalue $2(2k_R+1)^2$. The expression in the numerator
of the $l^2$ dependent term
is also of the same order. Its matrix elements can be calculated noting that its suffices to
neglect there the contribution of $l^2$, any contribution due to the minimal scale
is of higher order. Thus, the correction
is estimated as $l^2\otimes {\cal O}(1)$. 
The Lambshift is of the order of 1000Hz, corresponding
to about $\Delta E$ $\approx$ $4 \times 10^{-9}$eV, while the absolute value
of the energy of the ground state is about 13eV. Assuming the factor of
$l^2$ in (\ref{HyAt-4}) is of the order of 1GeV$^2$, we obtain
for the correction the the parenthesis in (\ref{HyAt-4})
$l^2 {\rm GeV}^2$ $\approx$ 
$\frac{\Delta E}{13}{\rm GeV}^2$ $\approx$ $3 \times 10^{-10}{\rm GeV}^{-2}$
~~~, 
which corresponds to a length parameter of about $1.7 \times 10^{-5}$GeV$^{-1}$, or 
$3.5 \times 10^{-6}$fm = $3.5 \times 10^{-19}$cm. In other words, in order that 
the effect of the length parameter
shows up in the measurement of the order of the Lamb-shift, it has to have this value.
Of course we do not see it yet, which leads to an upper estimate 
$l < 3.5 \times 10 ^{-19}~{\rm cm}$.

This is a more stringent restriction as obtained in
\cite{schuller2002}, within the Born-Infeld theory,
where an upper limit for the maximal
acceleration of ${\cal A}_m > 10^{22}\frac{{\rm m}}{{\rm sec}^2}$
was obtained. The minimal length is related to this maximal acceleration via
$l=\left({\cal A}_m\right)^{-1}$, which with $c=1$ translates to
an $l$ of $10^{-7}~{\rm cm}$ within the
Born-Infeld theory. 

\section{Conclusions}
\label{concl}

The motivation of this contribution was to obtain an estimation for the upper limit{
of a minimal length
in the context of the Hydrogen atom, using
pseudo-complex Quantum Mechanics, which provides this minimal
length. The formulation differs from \cite{pcQM} in that the explicit dynamical symmetry
$SO(4)$ of the hydrogen atom is exploited.
The space considered is flat and the 
definitions follow the one of \cite{Greiner-Muller}.

For $l=0$, the Hamiltonian of the Hydrogen atom depends
on a second order Casimir operator of $SO_x(4)$,
where the $x$ indicates that it is defined in the
coordinates space $x_k$ ($k=1,2,3,4$). When $l > 0$,
the corresponding operator in terms of $x_k$ and $p^x_k$
is no longer a Casimir operator, but can be expressed in
terms of the one in $SO_R(4)$, where $R$ refers to the pseudo-real space part. 

After having presented the essential mathematics
of the $SO_{pc}(4)$ structure and its relation to the
$(x_k,y_k)$ space, the Hamilton operator of the Hydrogen
atom was investigated, with the result that
the corrections to the level energies of the Hydrogen
atom is of the order of $l^2$.
Comparing to the Lambshift 
an upper limit of $l \leq 3.5 \times 10^{-19}~{\rm cm}$
was obtained, the best one up to now.

Finally, I want to stress that the use of a {\it length parameter}, 
which is unaffected by a Lorentz contraction, is of a quite
practical use, because Lorentz-invariance can be maintained,
simplifying a lot not only calculations but also the
investigation of the effects of a minimal length. This description
might be still useful as an effective description of the granulation
of space, in case it results that there is a minimal {\it physical}
length which violates Lorentz-invariance.

\section*{Acknowledgments}
We acknowledge financial support form DGAPA-PAPIIT (IN100421)

\section*{Appendix: Units}

The {\it natural units} are used, defined as: $\hbar = c = 1$.
In what follows, only {\it approximate} numerical values
for some constants are given, because exact values are not of interest here.
The condition 
$c=1$ $\approx$ $3\times 10^{8} \frac{{\rm m}}{{\rm sec}}$
tells us that time has units of length and its value is
$1{\rm sec} ~=~ 3 \times 10^{8}{\rm m}$.
From there, one obtains a relation of
length with energy: $1\hbar c$ $\approx$ 200MeVfm, namely
(1fm=$10^{-13}$cm)
$1{\rm fm} \approx 5 {\rm GeV}^{-1}$.

Using that 
$1eV \approx 1.6\times 10^{-19} \frac{{\rm kg}~{\rm m}^2}{{\rm sec}^2}$,
we obtain the kg in units of GeV:
$1{\rm kg} \approx 6 \times 10^{26}{\rm GeV}$.

For the pc-coordinates $X_\mu = x_\mu + lIy_\mu$ the units are
$[X_\mu ]=[x_\mu ]=[l]={\rm GeV}^{-1}$, $[y_\mu ]=1$
and for the pc-momenta $P_\mu = p_\mu + lIp^y_\mu$
$[P_\mu ]=[p_\mu]={\rm GeV}$, $[p^y_\mu]={\rm GeV}^2$.


\vskip 1cm
\begin{thebibliography}{00}

\bibitem{snyder1947} H. S. Snyder, Phys. Rev.
{\bf 71} (1947), 38.

\bibitem{moyal}  J. E. Moyal, Proc. Camb. Phil. Soc.
{\bf 45} (1949), 99.

\bibitem{string} E. Witten, Nucl. Phys. B {\bf 443}.
(1995), 85.

\bibitem{loop} T. Thiemann, Lect. Notes Phys.
{\bf 631} (2003), 41.

\bibitem{pcQM} P. O. Hess, Astronomische Nachrichten
{\bf 336} (2015), 739. 

\bibitem{Greiner-Muller}  W. Greiner and B. M\"uller,
{\it Quantum Mechanics: Symmetries}, (Springer,
Heidelberg, 1994).

\bibitem{book} P. O. Hess, M. Schäfer and W. Greiner, 
{\it Pseudo-Complex General Relativity}, (Springer, 
Heidelberg, 2015).

\bibitem{PPNP} P. O. Hess, Progr. Part. Nucl. Phys. 
{\bf 114} (2020), 103809.

\bibitem{kelly} P.F. Kelly, R.B. Mann, Classical Quantum Gravity{\bf  3} (1986) 705.

\bibitem{schuller2002} F. P. Schuller, Phys. Lett. B
{\bf 540} (1-2) (2002), 119.

\end{thebibliography}
\end{document}